\documentclass[structabstract]{aa}  
\usepackage{graphicx, amsmath, natbib}
\bibpunct{(}{)}{;}{a}{}{,}

\usepackage{txfonts}
\begin{document}
   \title{A hybrid moment equation approach to gas-grain chemical modeling}
   \titlerunning{}


   \author{Fujun Du
           \thanks{Member of the International Max Planck Research School
           (IMPRS) for Astronomy and Astrophysics at the Universities of Bonn
           and Cologne.}
           \and
           B\'ereng\`ere Parise
           }
   \authorrunning{F. Du \& B. Parise}

   \institute{Max-Planck-Institut f$\ddot{\rm u}$r Radioastronomie,
              Auf dem H$\ddot{\rm u}$gel 69, 53121 Bonn, Germany\\
              \email{fjdu@mpifr.de}
              }

   \date{}



  \abstract
   {In addition to gas phase reactions, the chemical processes on the surfaces
   of interstellar dust grains are important for the energy and material budget
   of the interstellar medium. The stochasticity of these processes requires
   special care in modeling. Previously methods based on the modified rate
   equation, the master equation, the moment equation, and Monte Carlo
   simulations have been used.}
   {We attempt to develop a systematic and efficient way to model the gas-grain
   chemistry with a large reaction network as accurately as possible.}
   {We present a hybrid moment equation approach which is a general
   and automatic method where the generating function is used to generate the
   moment equations. For large reaction networks, the moment equation is cut off
   at the second order, and a switch scheme is used when the average population
   of certain species reaches 1. For small networks, the third order moments
   can also be utilized to achieve a higher accuracy.}
   {For physical conditions in which the surface reactions are important, our
   method provides a major improvement over the rate equation approach, when
   benchmarked against the rigorous Monte Carlo results. For either very low or
   very high temperatures, or large grain radii, results from the rate equation
   are similar to those from our new approach. Our method is faster than the
   Monte Carlo approach, but slower than the rate equation approach.}
   {The hybrid moment equation approach with a cutoff and switch scheme is a
   very powerful way to solve gas-grain chemistry. It is applicable to large
   gas-grain networks, and is demonstrated to have a degree of accuracy high
   enough to be used for astrochemistry studies. Further work should be done to
   investigate how to cut off the hybrid moment equation selectively to make
   the computation faster, more accurate, and more stable, how to make the
   switch to rate equation more mathematically sound, and how to make the
   errors controllable. The layered structure of the grain mantle could also be
   incorporated into this approach, although a full implementation of the grain
   micro-physics appears to be difficult.}


   \maketitle


\section{Introduction}

The chemistry of the interstellar medium can be roughly divided into two types:
gas phase chemistry and grain surface chemistry. The two types of chemistry are
coupled by the adsorption and desorption processes.
Species adsorbed on the grain surface migrate in a random walk manner, and they
may react with each other upon encounter at the same site (a local potential
minimum). The products can be released back to the gas phase through certain
desorption mechanisms.  In addition to the gas phase chemistry, grain chemistry
is important for the material and energy budget of the interstellar medium. For
example, besides H$_2$, molecules such as methanol are believed to be formed on
the grain surfaces \citep{Garrod07}, because its relatively high abundance
\citep[see, e.g.,][]{Menten1988a} cannot be reproduced by gas phase chemistry.

Several methods have been used to model the gas-grain chemistry. In the rate
equation (RE) approach \citep[see, e.g., ][]{Hasegawa92}, the surface processes
are treated the same way as the gas phase processes. This works fine when the
number of reactants on a single grain is large (under the assumption that the
system is well stirred; see \citet{Gillespie07}), but might not be accurate
enough when the average populations\footnote{Here by ``population'' we mean the
number of a species in a volume of interest, and by ``average'' we mean an
ensemble average (i.e.  average over many different realizations of the same
system setup). Hence ``population'' can only take non-negative integer values,
while ``average population'' is a non-negative real number.} of some reactants
on a single grain is small. This failure of the rate equation is related to the
treatment of two-body reaction. For the REs to be applicable, the probability
of one reactant being present should be independent of that of another being
present.  This is not always true, especially when the average populations of
both reactants are low, in which case they might be highly correlated. The
flaws in employing the RE for grain-surface chemistry were pointed out by
\cite{Charnley97} and \cite{Tielens97}.

To remedy this problem, modification schemes based on some empirical,
heuristic, and/or physical reasoning have been applied to the RE approach
\citep{Caselli98, Stantcheva01}, and are called modified rate equation (MRE)
approach. The validity of this method has been questioned \citep{Rae03}. A
modification scheme developed by \citet{Garrod08} uses different functional
forms for different surface populations, taking various competition processes
and refinements into account. It has been shown to work very well, even for
very large reaction networks \citep{Garrod09}.

Mathematically, the gas-grain system should be viewed as a stochastic chemical
system \citep[see, e.g.,][]{McQuarrie67, Gillespie76, Charnley98a}, being
described by a probability distribution $P(\vec{x}, t)$, which is the
probability that the system has a population vector $\vec{x}$ at time $t$, with
$x_i$ being the number of the $i$th species in the system. The evolution
equation of $P(\vec{x}, t)$ is the so-called master equation, whose form is
determined by the reaction network.

Many sophisticated methods have been proposed (mainly outside the astro-chemical
community; see, e.g., the operator method described in \cite{Mattis98}, or the
variational approach used by \cite{Ohkubo08}) to solve the master equation.
However, these methods work fine only when either the chemical network is
small or some special assumptions are made in the derivation, thus their
validity in the general case should be questioned. It is unclear whether
these methods can be generalized to large complex networks.

The numerical solution of the master equation has also been performed
\citep{Biham01, Stantcheva02,Stantcheva04}. To limit the number of variables in
the set of differential equations and to separate the deterministic and
stochastic species, usually {\it a priori} knowledge of the system is required
in these studies. The steady state solution of the master equation
can also be obtained analytically in some very simple cases, such as the formation
of H$_2$ molecules on the grain surface \citep{Green01, Biham02}.

On the other hand, the master equation prescription can be ``realized'' through
a stochastic simulation algorithm (SSA), proposed by \cite{Gillespie76} (see
also \citet{Gillespie77} and \citet{Gillespie07}). In this approach, the
waiting time for the next reaction to occur, as well as which specific reaction
will occur are random variables that are completely determined by the master
equation, so this approach should be considered the most accurate. In principle,
multiple runs are needed to average out the random fluctuations, but in
practice this is unnecessary if one only cares about the abundant species. This
approach has been applied successfully to astrochemical problems
\citep{Charnley98a, Charnley01a, Vasyunin09}, even in the case of very large
networks \citep{Vasyunin09}. Besides providing results that are accurate, this
approach is very easy to implement.  However, it requires a very long run time
for large networks if a long evolution track is to be followed, although some
approximate accelerated methods do exist \citep[e.g.][]{Gillespie00}.

The SSA described above is somewhat different from a Monte Carlo (MC) approach
which has also been applied to astrochemistry \citep[e.g.][]{Tielens82};
however, this approach is not rigorously consistent with the master equation
\citep[see the comment by][]{Charnley05}, and can lead to a reaction
probability higher than 1 \citep{Awad05} in certain cases. The nomenclature of
these two approaches is not always consistent in the astrochemical
literature\footnote{For a discussion about the relations and differences
between ``stochastic simulation'' and ``Monte Carlo'', see \citet{Kalos08} and
\citet{Ripley08}.}. For example, the SSA used by \citet{Vasyunin09} is called
the Monte Carlo approach in their paper.  Hereafter, we use the term ``Monte
Carlo'' when referring to the rigorous stochastic simulation approach of
Gillespie.

By taking various moments of the master equation, the so-called moment equation
(ME) is obtained \citep{Lipshtat03, Barzel07a, Barzel07b}. This set of
equations describes the evolution of both the average population of each
species and the average value of the products of the population of a group of
species, usually cut off at the second order moments. Its formulation is
similar to that of the RE, so it is relatively easy to implement. Furthermore,
in this approach the gas phase chemistry and grain surface chemistry can be
coupled together naturally. It has been tested on small surface networks.

In the present paper, we propose yet another approach to modeling gas-grain
chemistry, named the hybrid moment equation (HME) approach. The goal is to find
a systematic, automatic, and fast way to modeling gas-grain chemistry as
accurately as possible. Our method is based on the ME approach. Different
approximations are applied to the MEs at different time depending on the
overall populations at that specific time. It is hybrid in the sense that the
RE and the ME are combined together. The basic modification and competition
scheme presented in \cite{Garrod08} can be viewed as a semi-steady-state
approximation to our approach (by assuming that the time derivatives of certain
second order moments are equal to zero), while our approach can also be viewed
as a combination of the ME approach of \cite{Barzel07a} and the RE. In our
approach, the MEs are generated automatically with the generating function
technique, and in principle MEs up to any order can be obtained this way. We
benchmark our approach against the exact MC approach (i.e. the SSA of
Gillespie).

The remaining part of this paper is organized as follows. In section
\ref{sec:DesHyb}, we review the chemical master equation and ME, then describe
the main steps of the HME approach. In section \ref{sec:Benchmark}, we
benchmark the HME approach with a cutoff at the second order and the RE
approach against the MC approach with a large gas-grain network; we also tested
the HME approach with a cutoff at the third order with a small network. In
section \ref{sec:Discussion}, we discuss the performance of the HME, and its
relation with previous approaches, as well as possibilities for additional
improvements. Our way of generating the MEs is described in Appendix
\ref{apdxA}. A surface chemical network we used for benchmark
is listed in Appendix \ref{apdxKeane}.


\section{Description of the hybrid moment equation (HME) approach}
\label{sec:DesHyb}

In this section, we first review both the chemical master and moment
equations. Although this content can be found in many other papers
\citep[e.g.,][]{Charnley98a, Gillespie07}, we present them here as they are the
basis of our HME approach. We then describe the MEs and REs for a simple set of
reactions as an example, to demonstrate how the HME approach naturally arise as
a combination of ME and RE. Finally we show the main steps of the HME
approach.

\subsection{The chemical master equation and the moment equation (ME)}

A chemical system at a given time $t$ can be described by a state vector
$\vec{x}$ which changes with time, with its $j$th component $x_j$ being the
number of the $j$th species in this system. As a chemical system is usually
stochastic, $\vec{x}$ should be viewed as a random variable, whose probability
distribution function $P(\vec{x}, t)$ evolves with time according to the master
equation \citep{Gillespie07}
\begin{equation} \partial_t P(\vec{x}, t) =
\sum_{i=1}^{M} [a_i(\vec{x}-\vec{\nu}_i) P(\vec{x}-\vec{\nu}_i, t) -
a_i(\vec{x}) P(\vec{x})], \label{eqn:mastereq}
\end{equation}
where $a_i(\vec{x})$ is called the propensity function, $a_i(\vec{x})\Delta
t$ is the probability that given a current state vector $\vec{x}$ an $i$th
reaction will happen in the next infinitesimal time interval $\Delta t$, and
$\vec{\nu}_i$ is the stoichiometry vector of the $i$th reaction.  The sum is
over all the reactions, and $M$ is the total number of reactions.

The ME is derived by taking moments of the master equation.
For example, for the \emph{first order} moment $\langle x_j\rangle$,
which is simply the average number of species $j$, $\langle x_j\rangle \equiv
\sum_{\vec{x}} P(\vec{x}, t) x_j$, its evolution is determined by
\citep{Gillespie07}
\begin{align}
   \partial_t\langle x_j\rangle=& \sum_{\vec{x}}\partial_t[P(\vec{x}, t)]x_j \nonumber\\
   =& \sum_i^M\sum_{\vec{x}} x_j[a_i(\vec{x}-\vec{\nu}_i) P(\vec{x}-\vec{\nu}_i, t) - a_i(\vec{x}) P(\vec{x})]\nonumber\\
   =& \sum_i^M\sum_{\vec{x}} [(x_j+\nu_{ij})a_i(\vec{x}) P(\vec{x}, t) - x_ja_i(\vec{x}) P(\vec{x})]\\
   =& \sum_i^M\sum_{\vec{x}} \nu_{ij}a_i(\vec{x}) P(\vec{x}, t)\nonumber
   = \sum_i^M \nu_{ij}\langle a_i(\vec{x})\rangle,\nonumber
\end{align}
where $\nu_{ij}$ is the $j$th component of the stoichiometry vector of the $i$th
reaction, i.e. the number of $j$th species produced (negative when being
consumed) by the $i$th reaction. For higher order moments, their corresponding
evolution equations can be similarly derived, although the final form will be
more complex. In Appendix \ref{apdxA}, we present another method based on the
generating function technique to derive the MEs, which is more suitable for
programming.

For the simplest network, in which all the reactions are single-body reactions,
$a_i(\vec{x})$ is a linear function of $\vec{x}$. In this case the ME is closed
and can easily be solved. However, when two-body reactions are present, this is
no longer true, as $\langle a_i(\vec{x})\rangle$ might be of a form $\langle
x_k (x_k-1) \rangle$ or $\langle x_k x_l \rangle$, which is of order two and
cannot be determined in general by the lower order moments. Hence additional
equations governing their evolution should be included, i.e., they should be
taken to be independent variables. The evolution equation of these second order
moments may also involve moments of order three, and this process continues
without an end, thus the ME is actually an infinite set of
coupled equations (although in principle they are not completely independent if
the chemical system being considered is finite, which leads to a
finite-dimensional space of state vectors).  The equation cannot be solved
without a compromise, e.g., a cutoff procedure, except for the simplest cases
in which an analytical solution is obtainable in the steady state.

\subsection{The MEs and REs for a set of reactions}

We take the following symbolic reactions as an illustrative example
\begin{align}
\mbox{Adsorption:}\quad & a \xrightarrow{k_{\rm ad}} A, \label{exareacaA} \\
\mbox{Evaporation:}\quad & A \xrightarrow{k_{\rm evap}} a, \\
\mbox{Surface reaction:}\quad & A + B \xrightarrow{k_{\rm AB}} C + D, \\
\mbox{Surface reaction:}\quad & A + A \xrightarrow{k_{\rm AA}} E, \label{exareacAA}
\end{align}
where the $k$s are the reaction rates of each reaction, A -- E are assumed to be
surface species that are distinct from each other, and ``a'' is the gas
phase counterpart of A.

In the following we first write down the MEs and REs for this system, then
discuss the relations and differences between them, as well as the relation
between a cutoff of MEs and a cutoff of master equations in previous studies.
These discussions will be essential to developing our HME approach.

\subsubsection{The MEs for this system}

The propensity functions for the above four reactions are $k_{\rm ad} a$,
$k_{\rm evap} A$, $k_{\rm AB} AB$, and $k_{\rm AA} A(A-1)$, respectively. Here
for convenience we use the letter ``A'' to represent both the name of a species
and the population of the corresponding species.

For the first order moments, we have
\begin{align}
\partial_t \langle A\rangle = & k_{\rm ad} \langle a\rangle - k_{\rm evap}
\langle A\rangle - k_{\rm AB} \langle A B\rangle - 2 k_{\rm AA} \langle A(A-1)
\rangle, \label{eqn:Aa} \\
\partial_t \langle C\rangle = &  k_{\rm AB} \langle A B\rangle, \label{eqn:Ca} \\
\partial_t \langle E\rangle = &  k_{\rm AA} \langle A(A-1)\rangle. \label{eqn:Ea}
\end{align}
Other similar equations are omitted.  The symbol $\langle *\rangle$
is used to represent the average population of ``*'' in the system; the average
should be understood as an ensemble average.
The second order moments $\langle A B\rangle$ and $\langle A(A-1)\rangle$ have
their own evolution equations, which are
\begin{align}
\partial_t \langle A B\rangle = & k_{\rm ad} \langle a B\rangle - k_{\rm evap}
\langle A B\rangle \notag \\ &- k_{\rm AB} [\langle A(A-1) B\rangle + \langle A
B(B-1)\rangle + \langle A B\rangle] \label{eqn:ABa} \\ & -2 k_{\rm AA} \langle
A(A-1)B\rangle, \notag
\end{align}
\begin{align}
\partial_t \langle A(A-1)\rangle = & 2k_{\rm ad} \langle a A\rangle - 2k_{\rm
evap} \langle A(A-1)\rangle \notag \\ &- 2k_{\rm AB} \langle A(A-1) B\rangle
\label{eqn:AAa} \\ & -2 k_{\rm AA} [2\langle A(A-1)(A-2)\rangle + \langle
A(A-1)\rangle]. \notag
\end{align}
For this simple example set of reactions (equation (\ref{exareacaA} --
\ref{exareacAA}) ), the above equations can be easily obtained from the master
equation \citep[see, e.g.,][page 8]{Lipshtat03}. In the general case (e.g.,
when A -- E are not completely distinct from each other), an automatic way of
obtaining the MEs is described in Appendix \ref{apdxA}. The method described there
is also applicable to moments with any order, and to all the common
reaction types in astrochemistry.

In general, the third order moments in the above equations cannot be expressed
as a function of the lower order moments, so they need their own differential
equations. In the case of a cutoff at the second order, the chain of equations,
however, stops here. We describe the method required to evaluate them in
section~\ref{sec:HMEapp}.

\subsubsection{The REs for this system}

When using REs, equations (\ref{eqn:Aa} -- \ref{eqn:AAa}) are replaced by
\begin{align}
\partial_t \langle A\rangle =\ & k_{\rm ad} \langle a\rangle - k_{\rm evap}
\langle A\rangle - k_{\rm AB} \langle A\rangle\langle B\rangle - 2 k_{\rm AA}
\langle A \rangle^2, \tag{\ref{eqn:Aa}$'$} \\
\partial_t \langle C\rangle =\ &  k_{\rm AB} \langle A\rangle\langle B\rangle,
\tag{\ref{eqn:Ca}$'$}\\
\partial_t \langle E\rangle =\ &  k_{\rm AA} \langle A\rangle^2.
\tag{\ref{eqn:Ea}$'$} \\
\partial_t [\langle A\rangle\langle B\rangle] =\ & k_{\rm ad} \langle a\rangle
\langle B\rangle - k_{\rm evap} \langle A\rangle\langle B\rangle \notag \\
&- k_{\rm AB} [\langle A \rangle^2\langle B\rangle + \langle A\rangle\langle
B\rangle^2] \tag{\ref{eqn:ABa}$'$} \\
& -2 k_{\rm AA} \langle A\rangle^2\langle B\rangle, \notag \\
\partial_t [\langle A\rangle^2] =\ & 2k_{\rm ad} \langle a\rangle\langle
A\rangle - 2k_{\rm evap} \langle A\rangle^2 \notag \\
&- 2k_{\rm AB} \langle A\rangle^2 \langle B\rangle -4 k_{\rm AA} \langle
A\rangle^3. \tag{\ref{eqn:AAa}$'$}
\end{align}
The equations for $\langle A\rangle$$\langle B\rangle$
and $\langle A\rangle^2$ are of course not needed in the RE approach but are
simply derived from equation (\ref{eqn:Aa}$'$) (and an omitted similar equation
for $\langle B\rangle$) using the chain rule of calculus.

\subsubsection{The relation between MEs and REs} \label{sec:relaMERE}
The differences between the MEs (equation \ref{eqn:Aa} -- \ref{eqn:AAa}) and
the REs (equation \ref{eqn:Aa}$'$ -- \ref{eqn:AAa}$'$) in the present case are as follows:
All the $\langle AB\rangle$ are replaced by $\langle A\rangle$$\langle
B\rangle$, all the $\langle A(A-1)\rangle$ are replaced by $\langle
A\rangle^2$, all the $\langle A(A-1)B\rangle$ are replaced by $\langle
A\rangle^2\langle B\rangle$, the $\langle AB(B-1)\rangle$ is replaced by
$\langle A\rangle\langle B\rangle^2$, and the $\langle A(A-1)(A-2)\rangle$ is
replaced by $\langle A\rangle^3$.  Furthermore, the term $k_{\rm AB}\langle
AB\rangle$ in equation (\ref{eqn:ABa}) and the term $k_{\rm AA}\langle
A(A-1)\rangle$ in equation (\ref{eqn:AAa}) disappear in the RE
(\ref{eqn:ABa}$'$) and (\ref{eqn:AAa}$'$).

These differences make clear why the REs are accurate when the involved species
are abundant (namely when $\langle A\rangle{\gg}1$ and $\langle B\rangle{\gg}1$).
This is because, in this case, $\langle AB\rangle$ can be approximated well by%
\footnote{Assuming Poisson statistics, we have
$$\frac{|\langle AB\rangle-\langle A\rangle\langle
B\rangle|}{\langle A\rangle\langle B\rangle}\lesssim \sqrt{\frac{1}{\langle
A\rangle}+\frac{1}{\langle B\rangle}}\ll1.$$}
$\langle A\rangle\langle B\rangle$, and $\langle A(A-1)\rangle$ can be
approximated well by $\langle A\rangle^2$.

The RE approach will be erroneous when $\langle A\rangle$ or $\langle B\rangle$
are smaller than 1 because, in this case, the correlation between $A$ and $B$
might cause $\langle AB\rangle$ to differ considerably from $\langle
A\rangle\langle B\rangle$, and the fluctuation in $A$ might cause $\langle
A(A-1)\rangle$ to differ considerably from $\langle A\rangle^2$. It can also be
viewed like this: in equation (\ref{eqn:ABa}$'$) and equation
(\ref{eqn:AAa}$'$) that govern the evolution of second order moments, the
omitted term $k_{\rm AB}\langle AB\rangle$  might be much larger than the
retained terms such as $\langle A\rangle^2\langle B\rangle$ or $\langle
A\rangle\langle B\rangle^2$, and the omitted term $k_{\rm AA}\langle
A(A-1)\rangle$ might be much larger than the retained term $\langle
A\rangle^3$.

\subsubsection{The relation between a cutoff of MEs and a cutoff of possible states in previous master equation approaches}\label{sec:relacutME}

In Eqs. (\ref{eqn:Aa} -- \ref{eqn:AAa}) we do not write terms such as $\langle
A(A-1)\rangle$ in the split form $\langle A^2\rangle - \langle A\rangle$.  We
keep terms such as $\langle A(A-1) B\rangle$ and $\langle A(A-1)(A-2)\rangle$
in their present forms intentionally. One reason for this is that terms such as
$\langle A(A-1)\rangle$ look more succinct and follow naturally from our way of
deriving them (see Appendix \ref{apdxA}). When $\langle A\rangle \gg 1$,
$\langle A(A-1)\rangle$ and $\langle AB\rangle$ can be directly replaced by
$\langle A\rangle^2$ and $\langle A\rangle\langle B\rangle$, respectively, to
obtain the RE formulation.

More importantly, this formulation can be directly connected to the cutoff
schemes in the previous master equation approaches \citep[e.g.,][]{Biham01,
Stantcheva02}. For example, in a scheme in which no more than two particles of A
are expected to be present on a single grain at the same time, we have $P(A{>}2)
= 0$. In this case, $\langle A(A-1)(A-2)\rangle = \sum_{A=3}^{\infty} P(A)
A(A-1)(A-2) = 0$. Thus we see that a cutoff at a population of two in the master
equation approach corresponds naturally to assigning a zero value to moments
containing $A$ more than twice, as far as the moments are defined in the form
presented above.

\subsection{The HME approach} \label{sec:HMEapp}

The HME approach is a combination of the ME and RE approaches. The basic idea
is that, for deterministic (average population $>$1) species, the REs are used,
while for stochastic (average population $<$1) species, the stochastic effects
are taken into account by including higher order moments in the equations.
Since a deterministic species may become stochastic as time goes by, and vice
versa, the set of ODEs governing the evolution of the system also changes with
time, and is determined dynamically. A flow chart of our HME code is shown in
Fig. \ref{fig:flowchart}.

We first set up all potentially needed MEs (using the procedure described
in Appendix~\ref{apdxA}), with a cutoff of moments at a prescribed order
(usually two). After this and some other initialization work, the program
enters the main loop.

The main loop contains an ODE solver because the system of MEs is a set of
ordinary differential equations (ODEs). We use the solver from the {\it
ODEPACK} package\footnote{Downloaded from www.netlib.org}.

Not all MEs and moments are used at all times; which ones are used is
determined dynamically. In each iteration of the main loop, we verify whether
some surface species have changed from stochastic to deterministic, or from
deterministic to stochastic. The gas phase species are always treated as
deterministic, regardless of how small their average populations are. In either
of these two cases, we re-examine all the moments, and determine the way to
treat them. There are four cases:
\begin{enumerate}
  \item All the first order moments are treated as independent variables.
  \item If a moment consists of only stochastic species, and its order is no
        larger than the prescribed highest allowed order, it will be treated as
        an independent variable, and the corresponding moment equation will be
        included and solved. For the sake of numerical stability, its value
        should be no larger than its deterministic counterpart. For example, if
        the ODE solver yields a value of $\langle AB\rangle>\langle
        A\rangle\langle B\rangle$, then the latter value will be assigned to
        $\langle AB\rangle$.
  \item If a moment consists of only stochastic species, and its order is
        larger than the prescribed highest allowed order, its value will be set
        to zero, and of course, its moment equation will not be solved.
        For example, if $\langle A\rangle{<}1$ and $\langle B\rangle{<}1$,
        then, with a highest allowed order set to two, moments such as $\langle
        A(A-1)(A-2)\rangle$ and $\langle A(A-1)B\rangle$ will be set to zero.
        This follows from the discussion in section \ref{sec:relacutME}.
  \item If a moment contains at least one deterministic species, it will not be
        treated as an independent variable, and its moment equation will not be
        solved. It can be evaluated in the following way: assuming that the
        moment under consideration has a form $\langle AB(B-1)\rangle$, and
        that $A$ is deterministic (i.e. $\langle A\rangle{>}1$), then the value
        of $\langle AB(B-1)\rangle$ is set to be $\langle A\rangle\langle
        B(B-1)\rangle$. If $B$ is also deterministic, then it will be evaluated
        as $\langle A\rangle\langle B\rangle^2$. This follows from the
        discussion in section \ref{sec:relaMERE}.
\end{enumerate}  
From these procedures, we see that the number of equations, as well as the form
of these equations will change when a transition between stochastic and
deterministic state of certain species occurs. Each time the ODE system is
updated, the ODE solver must therefore be re-initialized.

It seems possible to replace the sharp transition between the stochastic and
deterministic state of a species (based on whether its average population is
smaller than 1) with a smooth transition, e.g., using a weight function similar
to that in \citet{Garrod08}. However, it is not mathematically clear which
weight function we should choose, and an arbitrary one might cause some artificial
effects, so we prefer not to use this formulation.


\begin{figure}[htbp]
\centering
\includegraphics[width=0.5\textwidth, keepaspectratio]{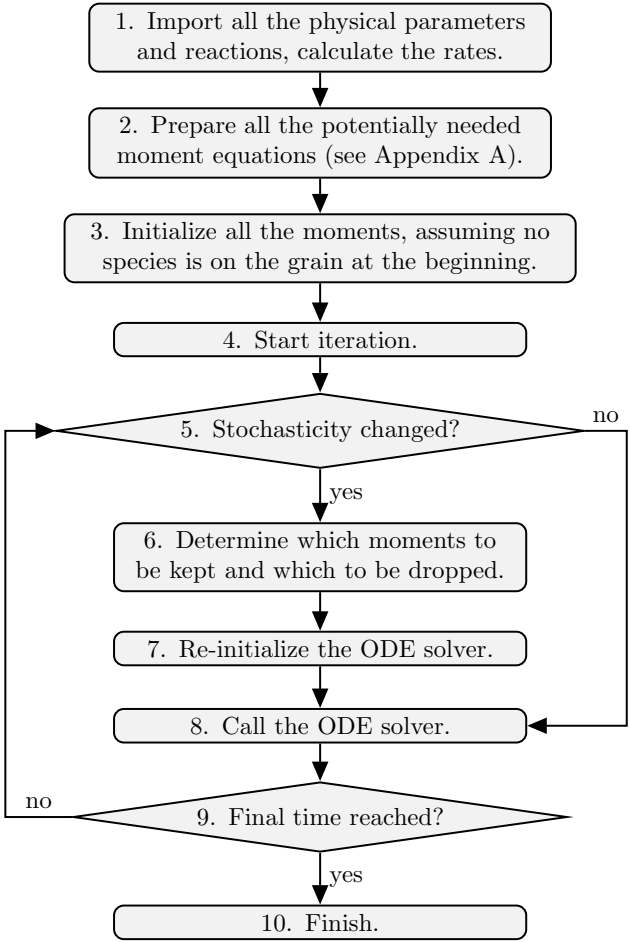}
\caption{A flow chart of the main components of our HME code. Steps 2, 5, 6, and
8 are described in detail in the text.}
\label{fig:flowchart}
\end{figure}

\section{Benchmark with the Monte Carlo approach} \label{sec:Benchmark}

We compare the results of our HME approach with those from the exact
stochastic simulation \citep{Gillespie07, Charnley98a, Charnley01a,
Vasyunin09}. The RE results are also compared for reference. As in previous
studies \citep{Charnley01a, Vasyunin09}, we consider a closed chemical system
in a volume containing exactly one grain particle. The number of each species
in this volume is called a ``population'', which can be translated into
an abundance relative to H nuclei by multiplying it by the dust-to-gas ratio,
which is $2.8\times10^{-12}(0.1\ \mu{\rm m}/r)^3$, where $r$ is the grain
radius, assuming an average molecular weight of 1.4, a dust-to-gas mass ratio
0.01, and a density of grain material of 2 g cm$^{-3}$.

In the MC approach, the number of each species in this volume at any time is
an integer. Owing to the large amount of time steps (${>}10^9$), it is
impractical to store all intermediate steps, so we average the population
of each species in time, weighted by the time intervals (remember that the
lengths of time intervals between reactions are also random in MC). Because of
this weighted average (rather than merely saving the state vector at certain
instants), the MC approach can resolve average populations much smaller than
one, although the fluctuations that are intrinsic to the MC approach can be
larger than the average populations when the latter is small.

We first demonstrate how our method works for a large gas-grain network.
We then show that for a small surface network, third order moments can also be
included to improve the accuracy.

\subsection{Test of the HME approach truncated at the second order on a large
gas-grain network} \label{sec:testlargeGG}

We use the ``dipole-enhanced'' version of the RATE06 gas phase reaction
network\footnote{http://www.udfa.net/}
\citep{Woodall07}, coupled with a surface network
of \citet{Keane97} (see Appendix \ref{apdxKeane}).
The surface network contains 44 reactions between
43 species, which is basically a reduced and slightly revised version of the
network of \citet{Tielens82}, containing the formation routes of the most
common grain mantle species, such as H$_2$O, CH$_3$OH, CH$_4$, NH$_3$, etc.
This surface network is not really large in comparison with some of the
previous works, such as that used by \citet{Garrod09}. However, it is already
essential for the most important species. The energy barriers for thermal
desorption and diffusion are taken from \citet{Stantcheva04}. Diffusion of H
atoms on the surface through quantum tunneling is included. Desorption by
cosmic rays is taken into account following the approach of
\citet{Hasegawa93a}. The rate coefficients of the gas phase reactions are
calculated according to \citet{Woodall07}, while the rate coefficients of the
surface reactions are calculated following \citet{Hasegawa92}. The initial
condition is the same as in \citet{Stantcheva04}.

We assumed a dust-to-gas mass ratio of 0.01. The grain mass density is taken to
be 2 g cm$^{-3}$, with a site density $5\times10^{13}$ cm$^{-2}$. Two grain
sizes have been used: 0.1 $\mu$m and 0.02 $\mu$m. A cosmic ray ionization rate
of $1.3\times10^{-17}$ s$^{-1}$ is adopted. Four different temperatures (10,
20, 30, 50 K) and three different densities ($10^3$, $10^4$, $10^5$ cm$^{-3}$)
have been used. In total, the comparison has been made for 24 different sets of
physical parameters. These conditions are commonly seen in translucent clouds
and cold dark clouds.

As in \citet{Garrod09}, we make a global comparison between the results
of MC, HME, and RE. For each set of physical parameters, the
comparisons are made at a time of $10^3$, $10^4$, and $10^5$ years. We
calculate the percentage of species for which the agreement between MC and
HME/RE is within a factor of 2 or 10. Only species with a population (either
from MC or from HME/RE) larger than 10 are included for comparison. This is
because for species with smaller populations, the intrinsic fluctuation in the
MC results can be significant. For several different sets of physical
parameters, we repeated the MC several times to get a feeling for how large
the fluctuation magnitude would be, although this is impractical for all the
cases.

\begin{figure*}[htbp]
\centering
\includegraphics[width=0.75\textwidth, keepaspectratio]{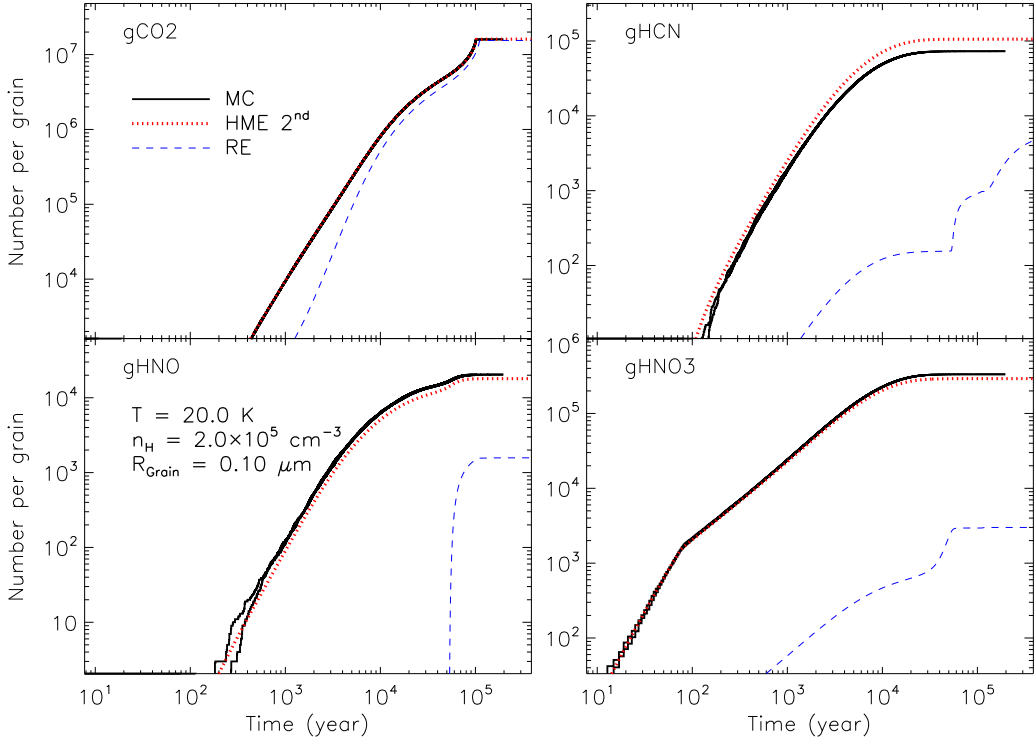}
\caption{Typical time evolution of the average populations of certain species from MC
(solid lines), HME to the 2$^{\rm nd}$ order (dotted lines), and RE
(dashed lines).  Note that the Monte Carlo has been repeated twice. The
y-axis is the number of each species in a volume containing exactly one grain.
To translate it into abundance relative to H nuclei, it should be multiplied by
$2.8\times10^{-12}$.  Physical parameters used: $T = 20$ K, $n = 2\times10^5$
cm$^{-3}$, grain radius = 0.1 $\mu$m.}
\label{fig:typicalEvol}
\end{figure*}

The comparison results are
shown in Table \ref{tab:agreeA} (grain radius = 0.1 $\mu$m) and Table
\ref{tab:agreeB} (grain radius = 0.02 $\mu$m). The HME approach always has a
better global agreement (or the same for several cases) than the RE approach in
the cases we tested. The typical time evolution of certain species is shown in Fig.
\ref{fig:typicalEvol}. In each panel of the figure, the species with a name
preceded with a ``g'' means it is a surface species.

\begin{figure*}[htbp]
\centering
\includegraphics[width=0.8\textwidth,
keepaspectratio]{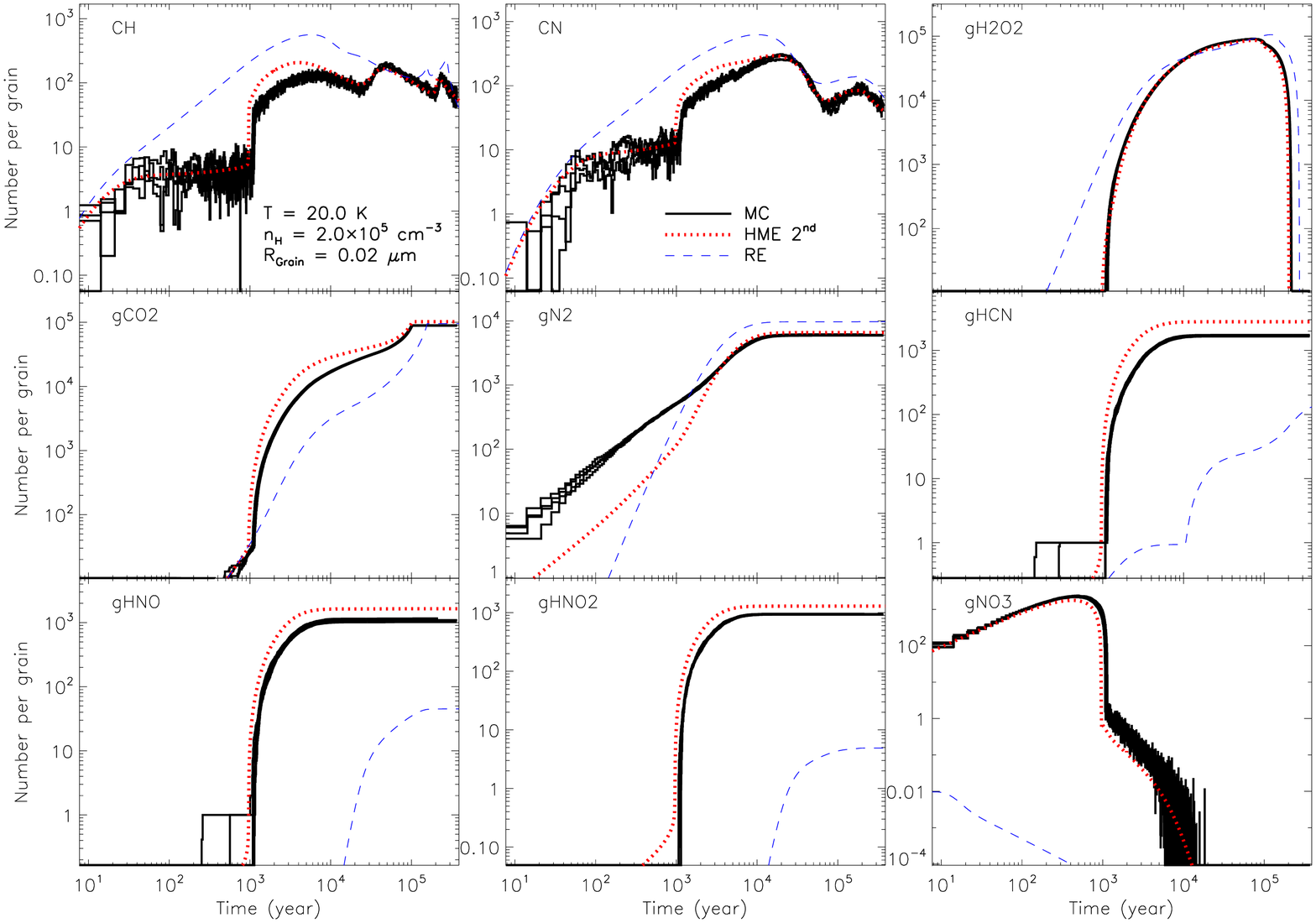}
\caption{Cases in which the agreement between the results of MC and
those of HME are not so good, especially at t = $10^3$ year. The
y-axis is the number of each species in a volume containing exactly one grain.
To translate it into abundance relative to H nuclei, it should be multiplied by
$3.5\times10^{-10}$. Physical parameters used: $T = 20$ K, $n = 2\times10^5$
cm$^{-3}$, grain radius = 0.02 $\mu$m.}
\label{fig:notSoGood}
\end{figure*}

The poorest agreement of HME (Fig. \ref{fig:notSoGood}) is at $t=10^3$ year for
T=20 K, $n_{\rm H}=10^5$ cm$^{-3}$, and grain radius = 0.02 $\mu$m. This is
mostly because at the time of comparison the populations of certain species
were changing very rapidly, so a slight mismatch in time leads to a large
discrepancy. This mismatch is probably caused by the truncation of higher-order
terms in the HME (see section \ref{sec:third}). For gN$_2$ in Fig.
\ref{fig:notSoGood}, its population seems to be systematically smaller in HME
than in MC during the early period, although the HME result matches the one
from MC at a later stage (after $3\times10^3$ years).

The RE is as effective as the HME in several cases, when the temperature is
either relatively low ($\sim$10 K) or high ($\sim50$ K) \citep[see
also][]{Vasyunin09}, and generally works better for a grain radius of 0.1
$\mu$m than of 0.02 $\mu$m. When the temperature is very
low, many surface reactions with barriers cannot happen (at least in the
considered timescales). On the other hand, when the temperature is high, the
surface species evaporate very quickly and the surface reactions are also
unable to occur. In these two extreme cases, the surface processes are
inactive, and the RE works fine.

The RE becomes problematic in the intermediate cases, when the temperature is
high enough for many surface reactions to occur, but not too high to
evaporate all the surface species; in these cases the HME represents a major
improvement over the RE. For a smaller grain radius, the population of each
species in a volume containing one grain will be smaller, thus the stochastic
effect will play a more important role, and the RE will tend to fail.

We note that, in the HME approach, there is no elemental
leakage except those caused by the finite precision of the computer. In
all the models that we have run, all the elements (including electric charge) are
conserved with a relative error smaller than $5{\times}10^{-14}$. The reason
why elemental conservation is always guaranteed is that either
the rate equations or the moment equations for the first order moments conserve
the elements.

\begin{table*}[htbp]
\caption{Percentage of agreement between the results from MC and those
from HME/RE. The comparison is only made between those
species with populations (from MC or HME/RE)
larger than 10.  The two numbers in each table entry means the percentage of
agreement within a factor of 2 or 10, respectively.  The grain radius is taken
to be 0.1 $\mu$m.}
\label{tab:agreeA}
\centering
\begin{tabular}{c c c c c c c c c c c c}
\hline\hline
\raisebox{1.5mm}{\phantom{I}} & \multicolumn{3}{c}{n$_{\rm H}$ = $2\times10^3$ cm$^{-3}$} & & \multicolumn{3}{c}{n$_{\rm H}$ = $2\times10^4$ cm$^{-3}$} & & \multicolumn{3}{c}{n$_{\rm H}$ = $2\times10^5$ cm$^{-3}$} \\ 
\cline{1-4} \cline{6-8} \cline{10-12}
\raisebox{1.5mm}{\phantom{I}}t &
$10^3$ yr & $10^4$ yr & $10^5$ yr & & $10^3$ yr & $10^4$ yr & $10^5$ yr & & $10^3$ yr & $10^4$ yr & $10^5$ yr \\
\hline
\multicolumn{12}{c}{hybrid moment equation} \\ \hline
T = 10 K & 100, 100  & 100, 100  & 100, 100 & & 100, 100 & 100, 100 &  97.6, 99.2 && 99.0, 100 & 100, 100 & 100, 100 \\
T = 20 K & 100, 100  & 100, 100  & 100, 100 & & 97.7, 98.9 & 98.2, 100 &  100, 100 && 100, 100 & 100, 100 & 100, 100 \\
T = 30 K & 100, 100  & 100, 100  & 100, 100 & & 98.8, 100 & 100, 100 & 99.2, 100 && 100, 100 & 100, 100 &  100, 100 \\
T = 50 K & 100, 100  & 100, 100  & 100, 100 & & 100, 100 & 100, 100 &  100, 100 && 100, 100 & 100, 100 & 97.9, 100 \\
\hline
\multicolumn{12}{c}{rate equation} \\ \hline
T = 10 K & 100, 100  & 100, 100  & 94.1, 98.8 & & 100, 100 & 100, 100 &  93.7, 98.4 && 99.0, 100 & 100, 100 & 95.8, 99.3 \\
T = 20 K & 90.2, 93.4  & 85.3, 90.7  & 83.6, 93.2 & & 95.5, 95.5 & 91.2, 95.6 &  92.3, 96.2 && 95.3, 96.2 & 95.0, 95.8 & 93.9, 96.6 \\
T = 30 K & 96.6, 96.6  & 95.5, 95.5  & 95.5, 97.0 & & 94.3, 96.6 & 98.1, 98.1 & 96.9, 97.7 && 94.9, 96.9 & 92.9, 97.4 & 40.0, 75.0 \\
T = 50 K & 100, 100  & 100, 100  & 100, 100 & & 100, 100 & 100, 100 &  100, 100 && 100, 100 & 100, 100 & 97.9, 100 \\
\hline\hline
\end{tabular}
\end{table*}

\begin{table*}[htbp]
\caption{Same as Table \ref{tab:agreeA} except a smaller grain radius of 0.02 $\mu$m is taken.}
\label{tab:agreeB}
\centering
\begin{tabular}{c c c c c c c c c c c c}
\hline\hline
\raisebox{1.5mm}{\phantom{I}} & \multicolumn{3}{c}{n$_{\rm H}$ = $2\times10^3$ cm$^{-3}$} & & \multicolumn{3}{c}{n$_{\rm H}$ = $2\times10^4$ cm$^{-3}$} & & \multicolumn{3}{c}{n$_{\rm H}$ = $2\times10^5$ cm$^{-3}$} \\ 
\cline{1-4} \cline{6-8} \cline{10-12}
\raisebox{1.5mm}{\phantom{I}}t &
$10^3$ yr & $10^4$ yr & $10^5$ yr & & $10^3$ yr & $10^4$ yr & $10^5$ yr & & $10^3$ yr & $10^4$ yr & $10^5$ yr \\
\hline
\multicolumn{12}{c}{hybrid moment equation} \\ \hline
T = 10 K & 100, 100  & 100, 100  & 100, 100 & & 100, 100 & 100, 100 & 100, 100 && 100, 100 & 100, 100 & 100, 100 \\
T = 20 K & 95.5, 100  & 100, 100  & 97.1, 100 & & 100, 100 & 100, 100 & 94.4, 100 && 73.0, 83.8 & 97.7, 100 & 98.3, 100 \\
T = 30 K & 100, 100  & 100, 100  & 100, 100 & & 100, 100 & 100, 100 & 100, 100 && 100, 100 & 100, 100 & 97.0, 100 \\
T = 50 K & 100, 100  & 100, 100  & 100, 100 & & 100, 100 & 100, 100 &  100, 100 && 100, 100 & 100, 100 & 100, 100 \\
\hline
\multicolumn{12}{c}{rate equation} \\ \hline
T = 10 K & 100, 100  & 100, 100  & 82.1, 94.9 & & 100, 100 & 100, 100 &  87.1, 95.2 && 100, 100 & 100, 100 & 94.8, 98.3 \\
T = 20 K & 87.0, 91.3  & 76.7, 83.3  & 71.4, 82.9 & & 74.3, 88.6 & 68.3, 87.8 & 28.6, 60.0 && 61.8, 82.4 & 42.3, 82.7 & 59.4, 92.2 \\
T = 30 K & 100, 100  & 95.7, 95.7  & 92.3, 92.3 & & 89.3, 92.9 & 90.9, 93.9 & 87.8, 90.2 && 82.1, 89.3 & 45.2, 90.3 & 24.3, 62.2 \\
T = 50 K & 100, 100  & 100, 100  & 100, 100 & & 100, 100 & 100, 100 &  100, 100 && 100, 100 & 100, 100 & 93.3, 93.3 \\
\hline\hline
\end{tabular}
\end{table*}

When comparing the results from the HME approach with those from MC
simulation, it is important to see how the intrinsic fluctuation in MC behaves.
If we assume the probability distribution of the population of a species, say
A, is Poissonian, then the variance of A is $\sigma^2(A)=\langle A\rangle$. Hence
if $\langle A\rangle$ is small, the relative fluctuation of the MC result can
be quite large. This fluctuation might be smoothed out by means of a weighted
average in time, but this procedure does not always work. This is why we choose
to only compare species with a population higher than 10, corresponding
to an abundance relative to H nuclei of $2.8\times10^{-11}$ (for grain radius =
0.1 $\mu$m) or $3.5\times10^{-9}$ (for grain radius = 0.02 $\mu$m). For a real
reaction network, it is usually difficult to predict the intrinsic fluctuation
in a MC simulation, unless it is repeated many times. These
fluctuations will not have any observational effects, because along a
line of sight there are always a large number of a certain species (as far as
it is detectable) and the fluctuations are averaged out.

We note that the gas phase processes are not treated identically in our HME
approach and MC simulation. In the MC approach, the gas phase processes are
always treated as being stochastic \citep[see, e.g.,][]{Charnley98a,
Vasyunin09}, in the same way as the surface processes. However, in our HME
approach, the gas phase species are treated in a deterministic way, i.e., REs
are always applied to them. This means that even if two reacting gas phase
species A and B both have average populations much smaller than one, we still
assume that $\langle AB\rangle = \langle A\rangle\langle B\rangle$.  This is
physically quite reasonable, because the presence of large amounts of reacting
partners in the gas phase (if not limited to a volume containing only one dust
grain; see, e.g., \citet{Charnley98a}) ensures that the RE is applicable.
However, although it might sound a bit pedantic, mathematically this is not
equivalent to the MC approach, and some discrepancies caused by this are
expected. For a large network, it is impractical to treat the gas phase
processes in the same way as the surface processes in the HME approach, because
in that case the number of independent variables in the ODE system will be
quite large (at least no less than the number of two-body reactions), and the
performance of the ODE solver will be degraded.


\subsection{Test of the HME approach truncated at the third order on a small
surface network} \label{sec:third}

To test the improvement in accuracy when the cutoff is made at a higher order,
we compare the results of the HME approach with a cutoff at the second order to
those obtained from the same approach with a cutoff at the third order.
We use a small surface reaction network
of \citet{Stantcheva04}, containing 17 surface reactions between 21 species,
producing H$_2$O, CH$_3$OH, CH$_4$, NH$_3$, and CO$_2$. No gas phase reactions
are included, except adsorption and desorption processes. The initial gas phase
abundances of the relevant species are obtained from the steady state solution
of the RATE06 network under the corresponding physical conditions.

\begin{figure*}[htbp]
\centering
\includegraphics[width=0.75\textwidth, keepaspectratio]{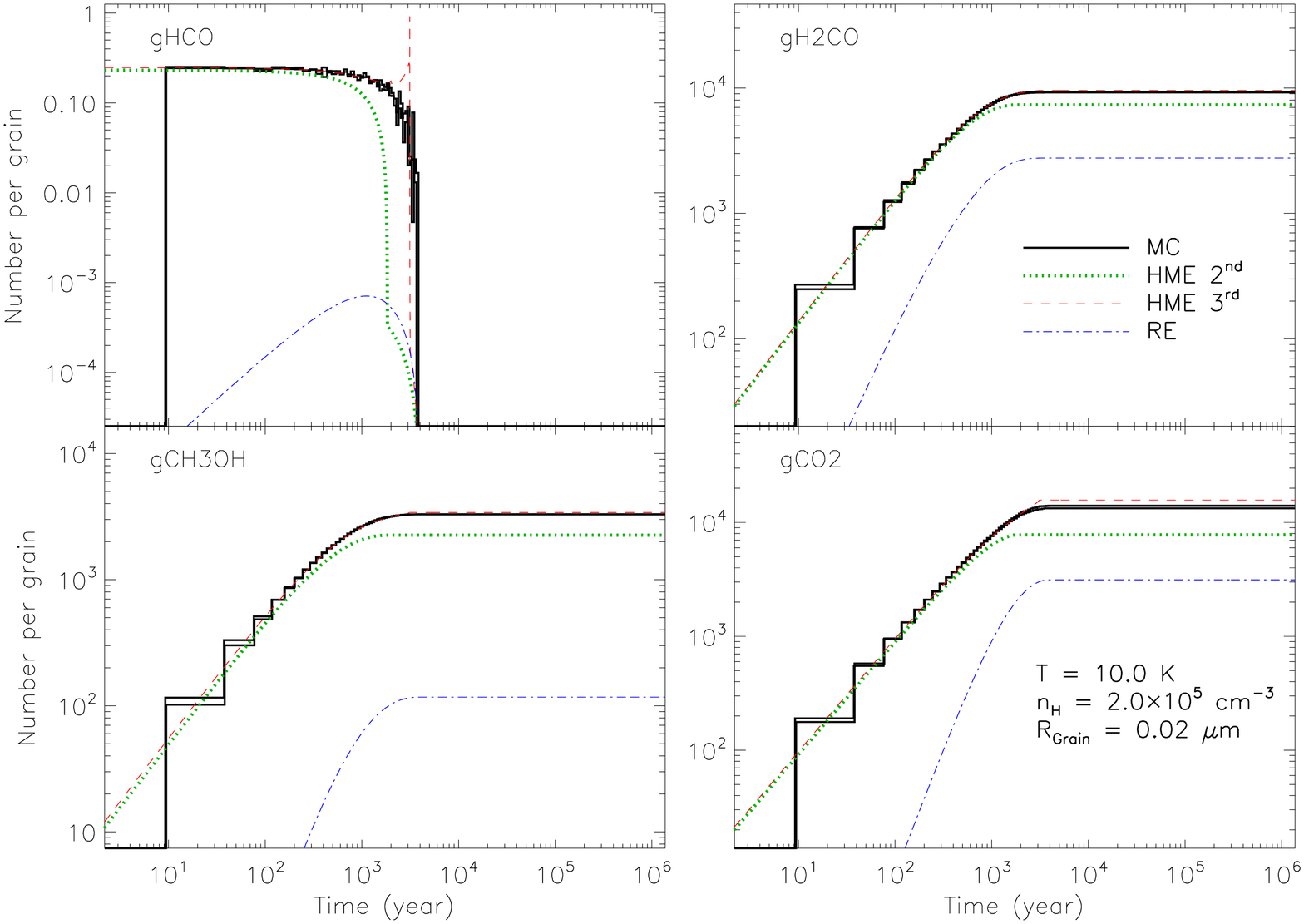}
\caption{Comparison of the results from MC, HME to the 2$^{\rm nd}$ order, HME
to the 3$^{\rm rd}$ order, and RE. The y-axis is the number of each species in
a volume containing exactly one grain.  To translate it into abundance relative
to H nuclei, it should be multiplied by $3.5\times10^{-10}$.  Physical
parameters used when running these models include $T = 10$ K, $n_{\rm H} =
2\times10^5$ cm$^{-3}$, grain radius = 0.02 $\mu$m.} \label{fig:thirdOrder}
\end{figure*}

As before, we run the HME, RE, and the MC code for different sets of physical
parameters. Although by transferring from the RE to the second order HME a
major improvement in accuracy can be obtained, the inclusion of the third
order moments to the HME usually only improves slightly over the second order
case.  In Fig. \ref{fig:thirdOrder}, we show an example (T = 10 K, $n_{\rm
H}=2\times10^5$ cm$^{-3}$, grain radius=0.02 $\mu$m), in which the distinctions
between the results from the second and third order HME are relatively large.

For several species, we note that the third order HME is still unable to match
the MC results perfectly, and for gHCO (Fig.  \ref{fig:thirdOrder}) the third
order HME even produces an artificial spike in the time evolution curve.  The
results from the third order HME are otherwise of greater accuracy than the
second order one, the abundances of gH$_2$CO and gCH$_3$OH in particular being
in almost perfect agreement with those from the MC approach. In the case of
gHCO, the timescale mismatch between HME and MC is alleviated by including the
third order moments.

It might be useful to see the difference between the second order HME and the
third order HME in a computational sense. For the current reaction network
with physical parameters described above, the number of variables (same as the
number of equations, which changes with time) is 145 initially in the second
order case, and this number becomes 705 for the third order case. To reach a
time span of $\sim$$10^6$ year, the second order HME takes about 3 seconds,
while the third order one takes about 220 seconds on a standard desktop
computer (a CPU @ 3.00 GHz with double cores, 4 GBytes memory). The number of
variables depends on the network structure, and it is not straightforward to derive
a formula to calculate it. Qualitatively, this number (as a function of the
number of reactions or the number of species) seems to increase with the cutoff
order less quickly than exponential growth. However, such a ``mild'' increase
affects the behavior of the ODE solver quite significantly. This is partly
because the solver contains operations (such as matrix inversion) that become
slower as the number of variables become larger. An increase in the number of
variables might also increase the stiffness of the problem, let alone the
memory limitation of the computer. For the larger reaction network described in
the previous section, the third order HME would involve about 5000 variables
and has not been tested successfully.


\section{Discussion} \label{sec:Discussion}

In our HME approach, we have used a general and automatic way to derive the
MEs. For large gas-grain networks, the MEs are cut off at the second order. For
small networks, a cutoff at the third order is possible and higher accuracy can
be obtained. We incorporate a switching scheme between the ME and RE when the
average population of a species reaches 1.

The results from HME are more accurate than those from the RE in the cases we
have tested, when benchmarked against the exact MC results. The abundances of
almost all the abundant species (${\gtrsim}2.8{\times}10^{-11}$ for a grain
radius of 0.1 $\mu$m and ${\gtrsim}3.5{\times}10^{-9}$ for a grain radius of
0.02 $\mu$m) from HME are accurate to within a factor of two, especially at
later stages of the chemical evolution, while in some cases nearly 40\% of the
results from RE are incorrect by a factor of at least ten.

In terms of computation time, our approach usually takes several tens of
minutes to reach a evolution time span of $10^6$ years, so it is slower than
the RE, but faster than the MC approach (which usually takes from several hours
to days). Our approach may also be slower than the MRE approach of
\citet{Garrod08}, because more variables (namely the moments with orders higher
than one) are present in our method, and the ODE system in HME is usually
stiffer. For example, for a moderate temperature many surface reactions can be
much faster than any gas phase reactions, and yield a very large coefficient in
some of the MEs. However, this is not the case in the stochastic regime of the
MRE approach, because when a competition scheme is used in MRE, such a large
coefficient does not appear. In this sense, we also advocate the MRE approach
of \citet{Garrod08}.

Mathematically, our approach is partially equivalent to the master equation
approach of \citet{Stantcheva04} in two respects.
1) They separated stochastic and deterministic species, which is similar to our
adopting RE for the abundant species.
2) They set a cutoff for the possible states of the stochastic species. This is
in essence equivalent to letting moments containing these species with order
higher than a certain number equal to zero.

Our approach can also be viewed as a combination of some of the ideas of
\citet{Garrod08} and \citet{Barzel07a}. The basic modification and competition
scheme in \citet{Garrod08} can be derived from the MEs, with a semi-steady
state assumption for the second order moments. \citet{Barzel07a} used the MEs,
but they did not include a switch scheme, and their way of deriving the MEs is
different from the one in the present paper.

There are still many possibilities for improvement. Although in principle
moments with any order can be included, the number of equations grows quite
quickly with the cutoff order, which makes the system of equations intractable
with a normal desktop computer. It is unclear whether it is possible to include
the moments selectively. It is unclear whether there are better, and more
mathematically well founded strategies than the switch at an average population
of 1. The present approach is usually stable numerically. However, this is not
always guaranteed, especially if higher order moments are to be included.  The
behavior of the numerical solution also depends on other factors, such as the
ODE solver being used and the tunable parameters for it, while the MC approach
does not have such issues. In this sense, the MC approach is the most robust.

Even in the accurate MC approach described above, the detailed morphology of
the grain surface and the detailed reaction mechanism is not taken into
account. One step in this direction would be to take into account the layered
structure of the grain mantle. This was done by \citet{Charnley01a} (see also
\citet{Charnley09}) by means of stochastic simulation.  It could also be
included in the HME approach, as far as the underlying physical mechanism could
be described by a master equation.

However, a microscopic MC approach has also been used to study the grain
chemistry \citep[see, e.g.,][]{Chang05, Cuppen07}. In this approach, the
morphology of the grain mantle and the interaction between species are modeled
in detail. As far as we know, this approach is only practical when the network
is small. It remains unclear whether it is possible to incorporate these
details into the current HME approach.

In some cases, errors caused by uncertainties in the reaction mechanism and
rate parameters might be larger than those introduced by the modeling method
\citep{Vasyunin08}. Hence, further experimental study and a more sophisticated
way of interpreting those results would be indispensable.



\begin{acknowledgements}
We thank Rob Garrod, Julia Roberts, Tom Millar, Guillaume Pineau des
For$\hat{\rm e}$ts, and Malcolm Walmsley for answering questions of the first
author during the early stage of this work. We also thank A.G.G.M. Tielens,
Anton Vasyunin, and Eric Herbst for discussions related to the present work,
and Antoine Gusdorf for useful comments.  This work is financially supported by
the Deutsche Forschungsgemeinschaft Emmy Noether program under grant
PA1692/1-1.  The first author acknowledges travel funding from the IMPRS for
astronomy and astrophysics at the Universities of Bonn and Cologne.
\end{acknowledgements}


\bibliographystyle{aa}
\bibliography{references}

\begin{appendix}

\section{A method to generate the moment equations based on the generating
function} \label{apdxA}

We describe our means of getting the MEs. Our method is
automatic, and can be easily coded into a computer program. It is applicable to
moments of any order and all the common astrochemical reactions. It makes use
of the probability generating function. While preparing the present paper, we
noted that \citet{Barzel10} also proposed a binomial
formulation of ME, which in essence is partly equivalent to our approach
presented here, although our way of deriving the MEs is quite
different from theirs.

For a probability distribution $P(\vec{x}, t)$, the corresponding generating
function is defined as \citep{vanKampen07}
\begin{equation} 
f(z_1, z_2, \ldots, t) = \sum_{\vec{x}} P(\vec{x}, t) z_1^{x_1} z_2^{x_2}
\cdots .
\end{equation}
Here all the $z_i$s should be thought of as merely symbols without any physical
meaning, and they have a one-to-one correspondence with the $x_i$s.

It is obvious that $f(\vec{z}=\vec{1}, t)\equiv1$, which is just the
normalization condition for probability. It is also easy to see that the average
population of
the $i$th species is
\begin{equation} 
\langle x_i\rangle = \partial_{z_i} f(z_1, z_2, \ldots, t) |_{\vec{z}=\vec{1}}.
\end{equation}
The right hand side of the above equation means taking the partial derivative
first, then assigning a value one to all the $z_i$s.

For the second order moment between two distinct species $i$ and $j$, we have
\begin{equation} 
\langle x_i x_j\rangle = \partial_{z_i}\partial_{z_j} f(z_1, z_2, \ldots, t)
|_{\vec{z}=\vec{1}}.
\end{equation}
If $i$ equals $j$ in the above equation, then what we actually get is
\begin{equation} 
\langle x_i (x_i-1)\rangle = \partial_{z_i}^2 f(z_1, z_2, \ldots, t)
|_{\vec{z}=\vec{1}}.
\end{equation}
In general, we have
\begin{equation} 
\langle x_i x_j x_k \cdots\rangle =
\partial_{z_i}\partial_{z_j}\partial_{z_k}\cdots f(z_1, z_2, \ldots, t)
|_{\vec{z}=\vec{1}}. \label{eqn:momentGen}
\end{equation}
If several of the subscripts are the same in the left hand side of the above
equation, say, $i=j=k$, then the second should be understood as ($x_j-1$),
while the third should be understood as ($x_k-2$), and so on.

From the master equation in Eq. (\ref{eqn:mastereq}), it seems possible to get an
equation for the evolution of $f(\vec{z}, t)$ in the general case. However, if
the propensity functions $a_i(\vec{x})$ (see Eq. (\ref{eqn:mastereq})) are
allowed to take any functional form, then this is not straightforward.
Fortunately, in practice $a_i(\vec{x})$ usually has a very simple form. On the
other hand, we note that in the right hand side of the master equation
in Eq. (\ref{eqn:mastereq}) the contributions from all the reactions are added
linearly. Hence the contribution of each reaction to the evolution of
generating function can be considered independently of each other.

We assume there is only one reaction in the network, which has a form
\begin{equation}
x_1 + x_2 + \cdots + x_n \xrightarrow{k} y_1 + y_2 + \cdots + y_m,
\label{eqn:reacxy}
\end{equation}
where the $x_i$s and $y_i$s represent the reactants and products, which do not
have to be different from each other. We also use these symbols to represent the
populations of the corresponding species. If, given a population of $x_1$,
$x_2$, \ldots, $x_n$, the probability that the above reaction will happen in a
unit time is $k x_1 x_2 \cdots x_n$ (namely, the propensity
function $a(\vec{x}) = k x_1 x_2 \cdots x_n$; the product should be understood
as explained in the sentence following Eq. (\ref{eqn:momentGen})), then the
generating function\footnote{Instead of using $z_i$s as symbols for the
independent variables of the generating function $f$, we use $x_i$s and $y_i$s
instead.  This will not cause any confusion.} will evolve according to
\begin{align}
\partial_t f
= k (y_1 y_2\cdots y_m - x_1 x_2\cdots x_n) \partial_{x_1} \partial_{x_2}\cdots
\partial_{x_n} f . \label{eqn:diffGen}
\end{align}
It is not difficult to derive the above equation from the master equation and
the definition of generating equation and our assumption about the propensity
function.

We note that equation (\ref{eqn:diffGen}) has a very simple pattern that is
easy to remember: (a) The constant coefficient is the rate coefficient; (b) In
the parenthesis, the \emph{symbols} of all the products are multiplied
together, with a coefficient $+1$, while the \emph{symbols} of all the
reactants are multiplied together, with a coefficient $-1$; (c) In the
differential part, all the reactants are present as they are in the left hand side
of equation (\ref{eqn:reacxy}), while none of the products appear.

We now attempt to derive the ME. We obtain the evolution equation of each moment
(as defined in Eq. (\ref{eqn:momentGen})) by simply differentiating both sides
of equation (\ref{eqn:diffGen}) with respect to the relevant components in the
moment, then setting all the symbols to a value of one.

For example, for the reaction $A+B\xrightarrow{k}C+D$, the evolution equation
of the generating function $f$ is
\begin{equation*}
\partial_t f = k (CD-AB)\partial_A\partial_B f.
\end{equation*}
For $\langle AB\rangle$, we differentiate both sides of the above equation by
$A$ and $B$. We obtain
\begin{align*}
\partial_t \partial_A\partial_B f = &k [(CD-AB)\partial_A^2\partial_B^2 f \\
& -A\partial_A^2\partial_B f - B \partial_A\partial_B^2 f - \partial_A\partial_B f].
\end{align*}
Next we assign a value of one to all the symbols (A -- D) appearing in the
resulting expressions, and ``translate'' the remaining terms into moments
(recalling the remark about equation (\ref{eqn:momentGen})), obtaining
\begin{align*}
\partial_t \langle AB\rangle = -k[\langle A(A-1)B\rangle+\langle
AB(B-1)\rangle+\langle AB\rangle].
\end{align*}

Although the above derivation involves differentiations, these operations can
be easily translated into some combinatorial rules and written as a computer
program. A recursive procedure is needed to generate all the potentially
needed moments up to a given order.

\Online
\pagebreak
\section{A surface reaction network we used to test our code}
\label{apdxKeane}
\begin{table}[htbp]
\caption{The surface network used in Section \ref{sec:testlargeGG} of this
paper. The original references listed below this table should be cited if these
data are to be used. We note that the validity of the numerical method (namely
the HME approach) presented in this paper does not depend on the specific test
network that we used.}
\label{tab:surfnetworkBig}
\centering
\begin{tabular}{l l l l l l l l r}
\hline\hline
Number  && \multicolumn{2}{c}{Reactants} && \multicolumn{2}{c}{Products} && $E_{{\rm reac}}$
(K) \\
\hline
1   && H        &  H       &&  H$_{2}$       &            &&       0.0  \\
2   && H        &  O       &&  OH       &            &&       0.0  \\
3   && H        &  OH      &&  H$_{2}$O      &            &&       0.0  \\
4   && H        &  O$_{2}$      &&  O$_{2}$H      &            &&    1200.0  \\
5   && H        &  O$_{2}$H     &&  H$_{2}$O$_{2}$     &            &&       0.0  \\
6   && H        &  H$_{2}$O$_{2}$    &&  H$_{2}$O      &  OH        &&    1400.0  \\
7   && H        &  O$_{3}$      &&  O$_{2}$       &  OH        &&     450.0  \\
8   && H        &  CO      &&  HCO      &            &&    1000.0  \\
9   && H        &  HCO     &&  H$_{2}$CO     &            &&       0.0  \\
10  && H        &  H$_{2}$CO    &&  CH$_{3}$O     &            &&    1500.0  \\
11  && H        &  H$_{2}$CO    &&  H$_{2}$COH    &            &&    1500.0  \\
12  && H        &  CH$_{3}$O    &&  CH$_{3}$OH    &            &&       0.0  \\
13  && H        &  H$_{2}$COH   &&  CH$_{3}$OH    &            &&       0.0  \\
14  && H        &  HCOO    &&  HCOOH    &            &&       0.0  \\
15  && H        &  N       &&  NH       &            &&       0.0  \\
16  && H        &  NH      &&  NH$_{2}$      &            &&       0.0  \\
17  && H        &  NH$_{2}$     &&  NH$_{3}$      &            &&       0.0  \\
18  && H        &  C       &&  CH       &            &&       0.0  \\
19  && H        &  CH      &&  CH$_{2}$      &            &&       0.0  \\
20  && H        &  CH$_{2}$     &&  CH$_{3}$      &            &&       0.0  \\
21  && H        &  CH$_{3}$     &&  CH$_{4}$      &            &&       0.0  \\
22  && H        &  CN      &&  HCN      &            &&       0.0  \\
23  && H        &  NO      &&  HNO      &            &&       0.0  \\
24  && H        &  NO$_{2}$     &&  HNO$_{2}$     &            &&       0.0  \\
25  && H        &  NO$_{3}$     &&  HNO$_{3}$     &            &&       0.0  \\
26  && H        &  NHCO    &&  NH$_{2}$CO    &            &&       0.0  \\
27  && H        &  NH$_{2}$CO   &&  NH$_{2}$CHO   &            &&       0.0  \\
28  && H        &  N$_{2}$H     &&  N$_{2}$H$_{2}$     &            &&       0.0  \\
29  && H        &  N$_{2}$H$_{2}$    &&  N$_{2}$H      &  H$_{2}$        &&     650.0  \\
30  && O        &  O       &&  O$_{2}$       &            &&       0.0  \\
31  && O        &  O$_{2}$      &&  O$_{3}$       &            &&    1200.0  \\
32  && O        &  CO      &&  CO$_{2}$      &            &&    1000.0  \\
33  && O        &  HCO     &&  HCOO     &            &&       0.0  \\
34  && O        &  N       &&  NO       &            &&       0.0  \\
35  && O        &  NO      &&  NO$_{2}$      &            &&       0.0  \\
36  && O        &  NO$_{2}$     &&  NO$_{3}$      &            &&       0.0  \\
37  && O        &  CN      &&  OCN      &            &&       0.0  \\
38  && C        &  N       &&  CN       &            &&       0.0  \\
39  && N        &  N       &&  N$_{2}$       &            &&       0.0  \\
40  && N        &  NH      &&  N$_{2}$H      &            &&       0.0  \\
41  && N        &  HCO     &&  NHCO     &            &&       0.0  \\
42  && H$_{2}$       &  OH      &&  H$_{2}$O      &  H         &&    2600.0  \\
43  && O        &  HCO     &&  CO$_{2}$      &  H         &&       0.0  \\
44  && OH       &  CO      &&  CO$_{2}$      &  H         &&      80.0  \\
\hline
\end{tabular}

\tablebib{
\citet{Keane97} and \citet{Tielens82}.
}
\end{table}
\end{appendix}

\end{document}